\PassOptionsToPackage{unicode}{hyperref}
\PassOptionsToPackage{hyphens}{url}
\documentclass[
]{article}
\usepackage{amsmath,amssymb}
\usepackage{iftex}
\ifPDFTeX
  \usepackage[T1]{fontenc}
  \usepackage[utf8]{inputenc}
  \usepackage{textcomp} % provide euro and other symbols
\else % if luatex or xetex
  \usepackage{unicode-math} % this also loads fontspec
  \defaultfontfeatures{Scale=MatchLowercase}
  \defaultfontfeatures[\rmfamily]{Ligatures=TeX,Scale=1}
\fi
\usepackage{lmodern}
\ifPDFTeX\else
\fi
\IfFileExists{upquote.sty}{\usepackage{upquote}}{}
\IfFileExists{microtype.sty}{% use microtype if available
  \usepackage[]{microtype}
  \UseMicrotypeSet[protrusion]{basicmath} % disable protrusion for tt fonts
}{}

\usepackage[backend=biber, style=apa]{biblatex} % Use biblatex for bibliography management
\bibliography{References}

\makeatletter
\@ifundefined{KOMAClassName}{% if non-KOMA class
  \IfFileExists{parskip.sty}{%
    \usepackage{parskip}
  }{% else
    \setlength{\parindent}{0pt}
    \setlength{\parskip}{6pt plus 2pt minus 1pt}}
}{% if KOMA class
  \KOMAoptions{parskip=half}}
\makeatother
\usepackage{xcolor}
\usepackage{graphicx}
\makeatletter
\def\maxwidth{\ifdim\Gin@nat@width>\linewidth\linewidth\else\Gin@nat@width\fi}
\def\maxheight{\ifdim\Gin@nat@height>\textheight\textheight\else\Gin@nat@height\fi}
\makeatother
\setkeys{Gin}{width=\maxwidth,height=\maxheight,keepaspectratio}
\makeatletter
\def\fps@figure{htbp}
\makeatother
\providecommand{\tightlist}{%
  \setlength{\itemsep}{0pt}\setlength{\parskip}{0pt}}
\ifLuaTeX
  \usepackage{selnolig}  % disable illegal ligatures
\fi
\IfFileExists{bookmark.sty}{\usepackage{bookmark}}{\usepackage{hyperref}}
\IfFileExists{xurl.sty}{\usepackage{xurl}}{} % add URL line breaks if available
\hypersetup{
  pdftitle={ClassiPyGRB: Machine Learning-Based Classification and Visualization of Gamma Ray Bursts using t-SNE},
  hidelinks,
  pdfcreator={LaTeX via pandoc}}

\title{ClassiPyGRB: Machine Learning-Based Classification and
Visualization of Gamma Ray Bursts using t-SNE}

\newcommand{\orcid}[1]{\href{https://orcid.org/#1}{\includegraphics[height=8pt]{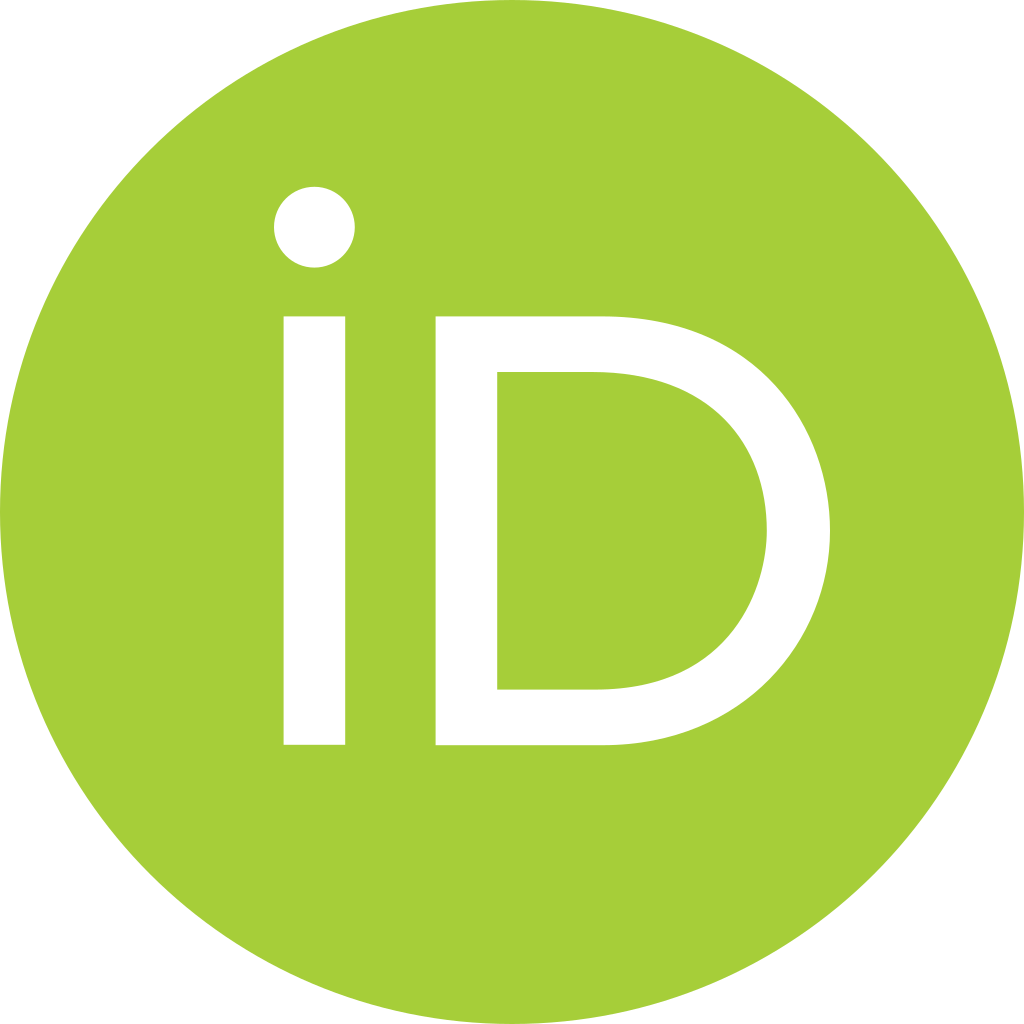}}}

\newcommand{\authorinfo}[4]{%
	#1\textsuperscript{#2} \orcid{#3} \\
	\vspace{5pt}
	\begin{minipage}{\linewidth}
		\centering
		\small #4
	\end{minipage}
}

\author{
	\authorinfo{Garcia-Cifuentes, Keneth}{1}{0009-0001-2607-6359}{
		$^1$Instituto de Ciencias Nucleares, Universidad Nacional Autónoma de México, Apartado Postal 70-264, 04510 México, CDMX, Mexico
	}
	\and
	\authorinfo{Rosa Leticia Becerra}{1,2}{0000-0002-0216-3415}{
		$^1$Instituto de Ciencias Nucleares, Universidad Nacional Autónoma de México, Apartado Postal 70-264, 04510 México, CDMX, Mexico\\$^2$Department of Physics, University of Rome - Tor Vergata, via della Ricerca Scientifica 1, 00100 Rome, IT
	}
	\and
	\authorinfo{Fabio De Colle}{1}{0000-0002-3137-4633}{
		$^1$Instituto de Ciencias Nucleares, Universidad Nacional Autónoma de México, Apartado Postal 70-264, 04510 México, CDMX, Mexico
	}
}
\date{22 May 2023 (Accepted April 8, 2024)}

\begin{document}
\maketitle

\begin{center}
	{\Large \LaTeX-based version of the paper. For the original manuscript, repository and software files, see the publication in {\color{blue}\href{https://doi.org/10.21105/joss.05923}{The Journal of Open Source Software}}.}
\end{center}

\section{Summary}\label{summary}

Gamma-ray burst (GRBs) are the brightest events in the universe. For
decades, astrophysicists have known about their cosmological nature.
Every year, space missions such as Fermi and SWIFT detect hundreds of
them. In spite of this large sample, GRBs show a complex taxonomy in the
first seconds after their appearance, which makes it very difficult to
find similarities between them using conventional techniques.

It is known that GRBs originate from the death of a massive star or from
the merger of two compact objects. GRB classification is typically based
on the duration of the burst (\cite{Kouveliotou:1993}). Nevertheless,
events such as GRB 211211A (\cite{Yang:2022}), whose duration of about 50
seconds lies in the group of long GRBs, has challenged this
categorization by the evidence of features related with the short GRB
population (the kilonova emission and the properties of its host
galaxy). Therefore, a classification based only on their gamma-ray
duration does not provide a completely reliable determination of the
progenitor.

Motivated by this problem, \cite{Jespersen:2020} and \cite{Steinhardt:2023} carried out analysis of GRB light curves by using the t-SNE algorithm, showing
that Swift/BAT GRBs database, consisting of light curves in four energy
bands (15-25 keV, 25-50 keV, 50-100 keV, 100-350 keV), clusters into two
groups corresponding with the typical long/short classification.
However, in this case, this classification is based on the information
provided by their gamma-ray emission light curves.

\textbf{ClassiPyGRB} is a Python 3 package to download, process,
visualize and classify GRBs database from the
\href{https://swift.gsfc.nasa.gov/about_swift/bat_desc.html}{Swift/BAT
Instrument} (up to July 2022). It is distributed over the GNU General
Public License Version 2 (1991). We also included a noise-reduction and
an interpolation tools for achieving a deeper analysis of the data.

\section{Statement of need}\label{statement-of-need}

The Swift Burst Alert Telescope (BAT) is a wide-field, hard Gamma-ray
detector that, since its launch in 2004, has played an important role
inthe high-energy astrophysical field. A preliminary query on the
Astrophysics Data System (ADS) with the keyword ``Swift/BAT'' indicates
that over 7000 research works have been uploaded to its platform (up to
December 2023). Furthermore, the number of studies per year is
increasingly, evidencing the relevance and impact of the use of the
Swift/BAT database.

Although the Swift/BAT database is publicly available, for first-time
users it might be a challenge to download and process the data. The data
are stored in multiple directories, depending on the GRB. Moreover, the
data are not pre-processed, and the user must perform the data
manipulation and interpolation themselves. These issues make the data
processing a time-consuming task. \texttt{ClassiPyGRB} is a Python 3
package that aims to solve these problems by providing a simple and
intuitive interface to download, process, visualize, and classify the
photometric data of GRBs from the Swift/BAT database.
\texttt{ClassiPyGRB} can also been used to promptly find similar GRBs
with some specific feature, such as a bright component
(\cite{Angulo-Valdez:2024}).

\begin{figure}
\centering
\includegraphics{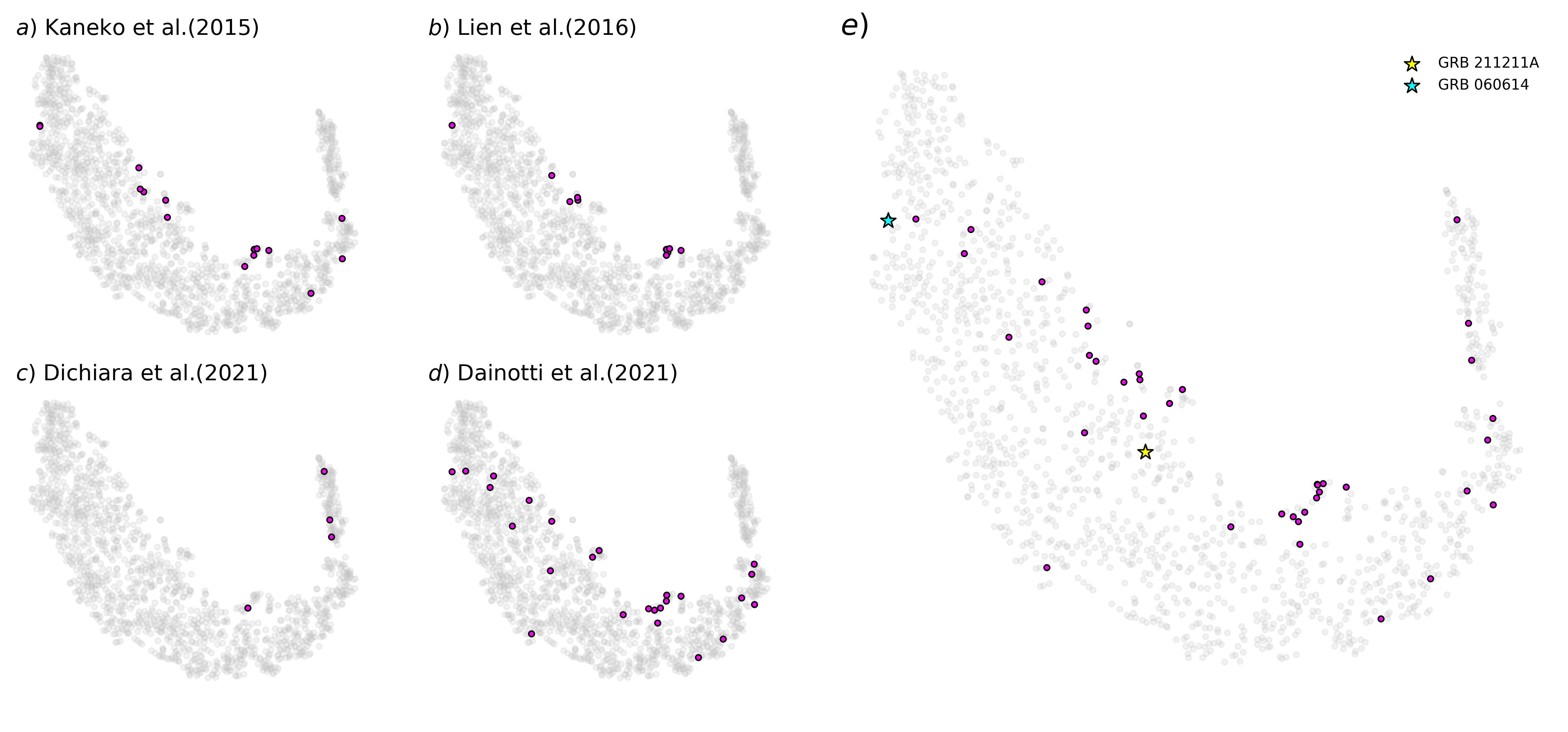}
\caption{t-SNE visualization map obtained for the noise-reduced dataset
binned at \(64\) ms with \(pp=30\). GRBs colored in magenta are
classified as Extended Emission by previous works. Image taken from
\cite{Garcia-Cifuentes:2023} \label{fig:fig1}}
\end{figure}

ClassiPyGRB allows researchers to query light curves for any GRB in the
Swift/BAT database simply and intuitively. The package also provides a
set of tools to preprocess the data, including noise/duration reduction
and interpolation. Moreover, The package also provides a set of
facilities and tutorials to classify GRBs based on their light curves,
following the method proposed by \cite{Jespersen:2020} and
\cite{Garcia-Cifuentes:2023} (see e.g., Figure 1). This method is based
on dimensionality reduction of the data using t-distributed Stochastic
Neighbour Embedding (t-SNE), where the user can visualize the results
using a Graphical User Interface (GUI). The user can also plot and
animate the results of the t-SNE analysis, allowing to perform a deeper
hyperparameter grid search. The package is distributed over the GNU
General Public Licence Version 2 (1991).

Although Swift/BAT offers basic functionality for data retrieval and
analysis, it is still necessary to implement a package that allows the
user to access and use the data in a simple way. In our case,
ClassiPyGRB is completely focused on GRB science. It complements what
other Python packages oriented to the Swift/BAT instrument do, such as:

\begin{enumerate}
\def\labelenumi{\arabic{enumi})}
\tightlist
\item
  The NITRATES pipeline (\cite{DeLaunay:2022}), designed for maximum
  likelihood-driven discovery and localization of Gamma-Ray Bursts.
\item
  BatAnalysis (\cite{Parsotan:2023}), a package specializing in photometry
  from all sources observed by the BAT instrument, not just in GRBs.
\end{enumerate}

Thus, ClassiPyGRB distinguishes itself by offering a comprehensive
solution that addresses the entire workflow of GRBs, from data retrieval
to classification. This package facilitates the entire process of GRB
analysis, ensuring accessibility, efficiency, and robustness for
researchers in the field of gamma-ray astronomy.

\section{Methodology and Structure of
ClassiPyGRB}\label{methodology-and-structure-of-classipygrb}

\texttt{ClassiPyGRB} mainly consists of three parts:

\begin{enumerate}
\def\labelenumi{\arabic{enumi})}
\tightlist
\item
  Retrieval and visualization of data from Swift/BAT: We implement an
  easy and fast code to download and plot an existing GRB post-processed
  data (e.g., Figure 2). There is the possibility to modify the time
  resolution of the data files (2ms, 8ms, 16ms, 64ms, 256ms, 1s and 10s)
  and analyze the data by selecting only some of the energy bands.
\end{enumerate}

\begin{figure}
\centering
\includegraphics{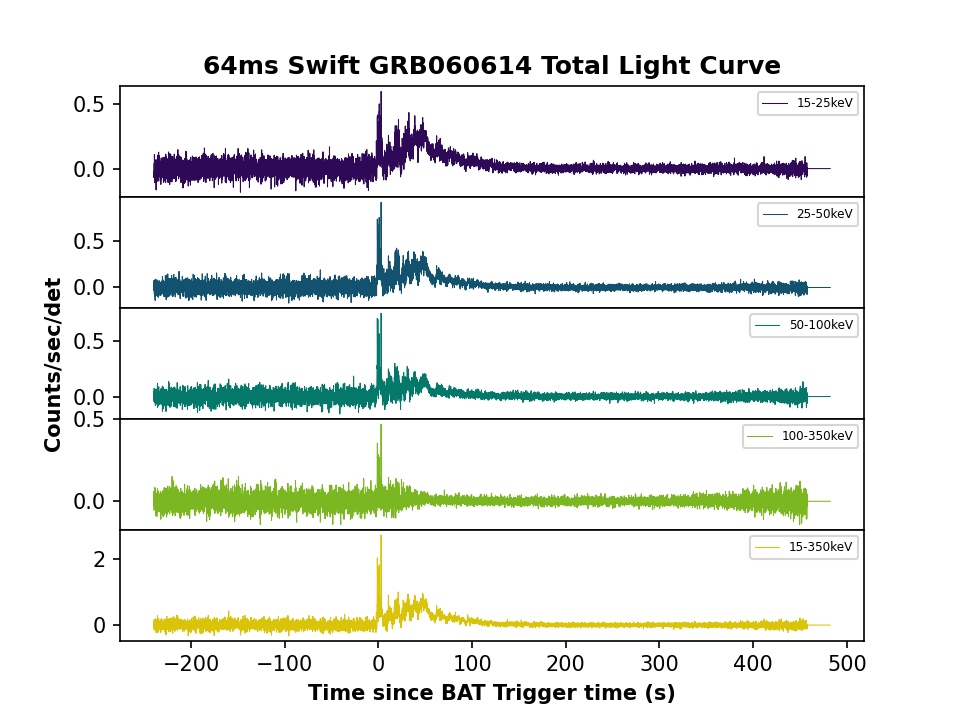}
\caption{Light curve of GRB 060614A. Image taken from
\cite{Garcia-Cifuentes:2023} \label{fig:fig2}}
\end{figure}

\begin{enumerate}
\def\labelenumi{\arabic{enumi})}
\setcounter{enumi}{1}
\tightlist
\item
  Data processing. \texttt{ClassiPyGRB} is able to:
\end{enumerate}

\begin{itemize}
\tightlist
\item
  Constrain the light curves by their duration \(T_\mathrm{100}\),
  \(T_\mathrm{90}\) or \(T_\mathrm{50}\).
\item
  Normalize the flux.
\item
  Standardize the temporal interval of all the sample (by zero-padding).
\item
  Improve the signal/noise (S/N) ratio applying the non-parametric noise
  reduction technique called FABADA \cite{Sanchez-Alarcon:2022} to each
  band for every single light curve.
\item
  Interpolate the flux between two specific times.
\end{itemize}

\begin{enumerate}
\def\labelenumi{\arabic{enumi})}
\setcounter{enumi}{2}
\tightlist
\item
  Visualization and plotting of t-SNE maps. \texttt{ClassiPyGRB}
  produces 2D visualization maps colored by the duration of GRBs.
\end{enumerate}

\begin{itemize}
\tightlist
\item
  It includes an intuitive graphic interface.
\item
  It is possible to add either of the two features to the t-SNE maps or
  to visualize the raw data.
\item
  Manipulation of the t-SNE parameters.
\item
  Visualization of the light curves from the events with and without
  noise-reduction.
\item
  There is the possibility of working only on the desired bands of
  Swift/BAT.
\item
  Specific events can be searched for and highlighted on the display.
\end{itemize}

Any plots created with \texttt{ClassiPyGRB} can be customized by the
user.

Note: Algorithms such as t-SNE visualization maps are very sensitive to
any change in the perplexity and learning rate parameters. Therefore, as
is the case when using any of these visualization techniques derived
from machine learning, care must be taken in the correct interpretation
of the data.

This repository requires Python 3.8 or higher, and the necessary
packages will be automatically handled during installation. Other
packages will be required optionally in Documentation (i.e., Jupyter).

\section{Acknowledgements}\label{acknowledgements}

KSGC acknowledges support from CONAHCyT fellowship. RLB acknowledges
support from CONAHCyT postdoctoral fellowship.

\printbibliography

\end{document}